\documentclass[prb,reprint,aps,superscriptaddress,longbibliography]{revtex4-2}
\usepackage{graphicx}% Include figure files
\usepackage{dcolumn}% Align table columns on decimal point
\usepackage{bm}% bold math
\usepackage{braket}
\usepackage{amsmath}
\usepackage{amssymb}
\usepackage{feynmp}
\usepackage{graphics}
\usepackage{enumerate}
\usepackage[dvipsnames]{xcolor}
\usepackage{comment}
\usepackage[normalem]{ulem} % sout 
\usepackage[hidelinks]{hyperref}% add hypertext capabilities

\begin{document}

\title{Pressure--enhanced fractional Chern insulators in moir\'e transition metal dichalcogenides along a magic line}

\author{Nicol\'as Morales-Dur\'an}
\email{na.morales92@utexas.edu}
\affiliation{Department of Physics, The University of Texas at Austin,  Austin, TX 78712, USA}
\affiliation{Center for Computational Quantum Physics, Flatiron Institute, 162 5th Avenue, New York, New York 10010, USA}

\author{Jie Wang}
\email{jiewang.phy@gmail.com}
\affiliation{Center for Mathematical Sciences and Applications, Harvard University, Cambridge, MA 02138, USA}
\affiliation{Department of Physics, Harvard University, Cambridge, MA 02138, USA}

\author{Gabriel R. Schleder}
\affiliation{John A. Paulson School of Engineering and Applied Sciences,
Harvard University, Cambridge, Massachusetts 02138, USA}
\affiliation{Brazilian Nanotechnology National Laboratory (LNNano), CNPEM, 13083-970 Campinas, São Paulo, Brazil}

\author{Mattia Angeli}
\affiliation{John A. Paulson School of Engineering and Applied Sciences, Harvard University, Cambridge, Massachusetts 02138, USA}

\author{Ziyan Zhu}
\affiliation{Stanford Institute for Materials and Energy
Sciences, SLAC National Accelerator Laboratory, Menlo Park, CA 94025, USA}

\author{Efthimios Kaxiras}
\affiliation{Department of Physics, Harvard University, Cambridge, MA 02138, USA}
\affiliation{John A. Paulson School of Engineering and Applied Sciences, Harvard University, Cambridge, Massachusetts 02138, USA}

\author{Cécile Repellin}
\email{cecile.repellin@lpmmc.cnrs.fr}
\affiliation{Univ. Grenoble Alpes, CNRS, LPMMC, 38000 Grenoble, France}

\author{Jennifer Cano}
\email{jennifer.cano@stonybrook.edu}
\affiliation{Department of Physics and Astronomy, Stony Brook University, Stony Brook, New York 11794, USA}
\affiliation{Center for Computational Quantum Physics, Flatiron Institute, 162 5th Avenue, New York, New York 10010, USA}

\begin{abstract}
We show that pressure applied to twisted WSe$_2$ can enhance the many-body gap and region of stability of a fractional Chern insulator at filling $\nu = 1/3$. Our results are based on exact diagonalization of a continuum model, whose pressure-dependence is obtained through {\it ab initio} methods.
We interpret our results in terms of a {\it magic line} in the pressure-{\it vs}-twist angle phase diagram: along the magic line, the bandwidth of the topmost moir\'e valence band is minimized while simultaneously its quantum geometry nearly resembles that of an ideal Chern band.
We expect our results to generalize to other twisted transition metal dichalcogenide homobilayers.
\end{abstract} 

\maketitle

{\em Introduction --- }
Moir\'e materials provide a highly tunable platform where many quantum phases of matter can be predicted, simulated, and explored.
Of particular interest is the potential to realize a fractional Chern insulator (FCI), the zero field analog of the fractional quantum Hall (FQH) effect \cite{FQH}.
FCIs appear in a number of lattice models with fractionally filled flat Chern bands \cite{Wen_FCI,Neupert_FCI,DasSarma_Sun_FCI,Sheng_FCI,Bernevig_Regnault_FCI,Review_ParameswaranRoySondhi,Review_BergholtzLiu,Review_BergholtzLiu2}. While these early models lacked a clear electronic realization, the unique interplay between topology and interactions in moir\'e materials brings the experimental realization of FCIs within reach. In fact, recently FQH-like states in magic angle twisted bilayer graphene (TBG) have been observed under reasonably small magnetic fields ($B \sim 5$T) \cite{Experiment_FCITBG}. 
But as the field is reduced, the finite-field ground state becomes unstable to a charge density wave (CDW), consistent with the close competition between the FCI and CDW phases found in theoretical studies \cite{Lauchli_FCI,Bergholtz_FCI1,Repellin_FCI_TBG,Ashvin_FCI1,Ashvin_FCI2,JunkaiJWLF22}. 

Thus, an experimental realization of the long-sought fractional quantum Hall physics at zero field remains to be found. Twisted transition metal dichalcogenides (TMDs) offer an alternative platform to TBG. Chern bands were predicted in certain TMD homo- \cite{FengchengTopology} and hetero-bilayers \cite{LiangFuTextures} and experimentally confirmed in MoTe$_2$/WSe$_2$ \cite{Cornell_QAH1,Cornell_QAH2,Cornell_QAH3}. Subsequent numerical studies predicted that at filling $\nu=1/3$ of the topmost moiré band, twisted MoTe$_2$ \cite{KaiSunFCI} and WSe$_2$ \cite{LiangFuFCI} homobilayers could host FCI ground states at zero magnetic fields. 

In this work, we show that pressure applied to twisted WSe$_2$ provides a new tuning knob to enhance the stability of the FCI phase. Specifically, we introduce a region in the pressure-{\it vs}-twist angle phase diagram where the FCI indicators \cite{Roy_Ideal,ParameswaranRoy_Ideal,Ideal_Jie,LedwithFCI,positionmomentumduality,kahlerband1,kahlerband2,kahlerband3,vortexability22} -- namely bandwidth, Berry curvature fluctuations, and trace of the quantum metric --  are near simultaneously optimized. We then use exact diagonalization to demonstrate an FCI ground state at $\nu=1/3$ stabilized in the region of near-ideal band geometry, whose many-body gap increases with pressure. We further address the competition between this FCI state and a CDW. Though our calculations are specific to twisted WSe$_2$, we expect similar features to appear in other twisted TMDs where flat topological bands emerge. 
 
{\em Moiré TMD topological bands under pressure--- }The low-energy physics of twisted homobilayers such as WSe$_2$ or MoTe$_2$ can be accurately described by a continuum model. The most general valley-projected Hamiltonian is given, in layer space, by \cite{FengchengHubbard,FengchengTopology}
\begin{align}
    \mathcal{H}^K=
    \begin{pmatrix}
    h^b({\bm k})&T({\bm r})\\
    T^{\dagger}({\bm r})&h^t({\bm k})-E_{\text{off}}
    \end{pmatrix}.
    \label{BMHamiltonian}
\end{align}
Due to spin-valley locking, carriers have only one pseudospin degree of freedom. The Hamiltonian describing the other valley is related by time-reversal symmetry. The diagonal terms of Eq. \eqref{BMHamiltonian} describe carriers populating the topmost valence band of either the bottom ($b$) or top  ($t$) layer. They consist of the quadratic dispersion of a single TMD layer folded into the moiré Brillouin zone, plus the moir\'e potential $\Delta^{b/t}({\bf r})$ due to the presence of the other layer, i.e.,
\begin{align}
    h^{b/t}({\bm k})&=-\frac{\hbar^2}{2m^*}{\bm k}^2+\Delta^{b/t}({\bm r}), \\
    \Delta^{b/t}({\bm r})=&2V_m\sum_{j=1,3,5} \cos\left({\bm b}_j\cdot {\bm r} \pm \psi \right),
    \label{Moirepotential}
\end{align}
where the vectors ${\bf b}_j=4\pi/\sqrt{3}a_M(\cos(\pi j/3),\sin(\pi j/3))$ belong to the first shell of reciprocal lattice vectors, with $a_M$ the moir\'e length, $m^*$ is the effective carrier mass, and $V_m$ and $\psi$ are parameters that determine the strength and spatial pattern of the moir\'e potential, respectively. The interlayer tunneling in Eq. \eqref{BMHamiltonian} is given by
\begin{align}
    T({\bf r})=\omega\left(1+e^{i\frac{2\pi}{3}\alpha}e^{i {\bf b}_2\cdot {\bf r}}+e^{i\frac{2\pi}{3}2\alpha}e^{i {\bf b}_3\cdot {\bf r}}\right),
\end{align}
where $\omega$ is the tunneling amplitude, $\alpha=0$ for AA--stacking ($0^{\circ}$ rotation between layers, as considered in this paper) and $\alpha=1$ for AB--stacking ($180^{\circ}$ rotation between layers) \cite{LiangFuMagic}. Finally, $E_{\text{off}}$ describes the offset between the two topmost bands from each layer, which vanishes for homobilayers in the absence of a displacement field, i.e., an interlayer potential difference. The effect of displacement field has been studied \cite{LiangFuFCI,ColumbiaCano} and tends to make bands more dispersive, disfavouring FCI stabilization \cite{Cornell_QAH1,Cornell_QAH2}; henceforth we set $E_{\text{off}}=0$.

\begin{figure}[t!]
\centering
\includegraphics[width=0.95\linewidth]{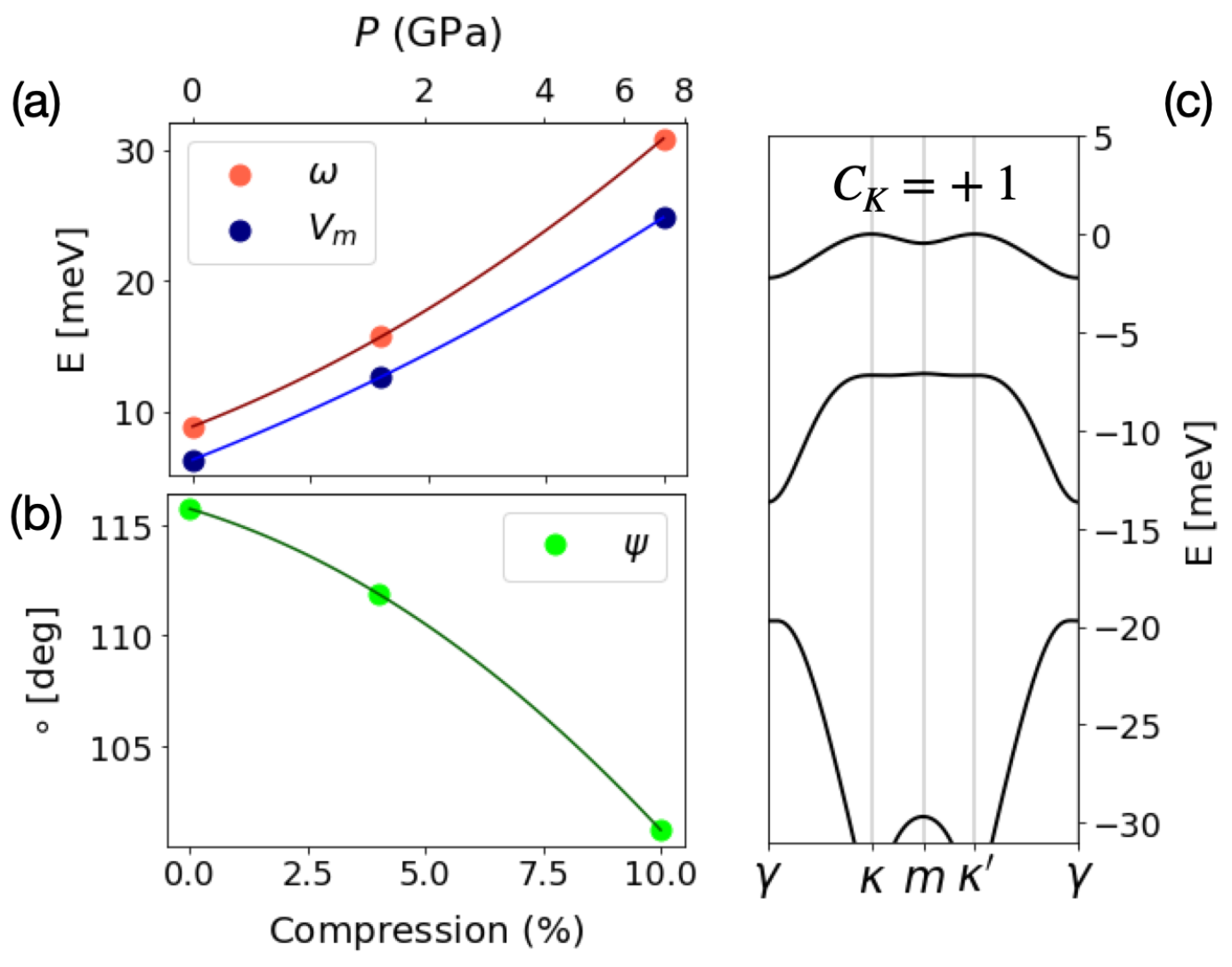}
\caption{(a)-(b) Evolution of continuum model parameters $V_m$, $\omega$ and $\psi$ for WSe$_2$ as a function of compression percentage (bottom axis) and applied pressure (top axis). (c) $K$-valley-projected bandstructure, corresponding to $P=2$ GPa and for $\theta=1.8^{\circ}$. The topmost moiré valence band from valley $K (K^{\prime})$ has Chern number $C_K=+1~(C_{K^{\prime}}=-1)$.}
\label{fig:BandstructureParameters}
\end{figure}

We focus on twisted AA-stacked WSe$_2$. Using {\it ab initio} calculations \cite{Supplemental}, we obtain an effective mass $m^* = 0.337\,m_0$ and compute the continuum model parameters ($V_m, \psi, \omega$) over a range of experimentally achievable pressures, $P$, or equivalently, sample compression percentages. The behaviors of $V_m$ and $\omega$ as a function of sample compression and pressure are shown in Fig. \ref{fig:BandstructureParameters}(a), while the corresponding evolution of the phase $\psi$ is shown in  Fig. \ref{fig:BandstructureParameters}(b). Both the tunneling amplitude $\omega$ and moiré strength $V_m$ increase quadratically with compression percentage, supporting the intuition that applied pressure pushes the layers closer together, increasing both the interlayer tunneling and the moiré potential. A similar trend was predicted in TBG \cite{PressureTBG}, followed by experimental realization \cite{CoryPressure1,CoryPressure2}. Thus, we expect the same tendency to hold for other twisted TMD homobilayers \cite{Supplemental}. Fig.~\ref{fig:BandstructureParameters}(c) shows an example of the obtained bandstructure at $P=2$ GPa. For the parameter range considered in this work, the topmost moiré valence band is always topological with Chern number $C_{K/K^{\prime}}=\pm 1$, consistent with previous calculations at $P=0$ \cite{FengchengTopology,LiangFuMagic}. 
Time-reversal symmetry enforces that the Chern numbers at valley $K$ and $K^{\prime}$ are opposite.

{\em FCI indicators --- }
Comparing a Chern band to the lowest Landau level (LLL) yields single-particle indicators of the stability of a putative FCI phase when the Chern band is partially filled \cite{Review_ParameswaranRoySondhi, Review_BergholtzLiu}.
Specifically, the LLL has vanishing bandwidth, homogeneous Berry curvature, and an ideal quantum geometry \cite{Roy_Ideal,ParameswaranRoy_Ideal,Moller_Ideal, Ideal_Jie,LedwithFCI,Review_BergholtzLiu2,BauerGeometry2016, simonGeometry2020}.
Though the quantum geometry has been an intense subject of recent study, particularly in relation to TBG, it has not been studied in moiré TMDs.

\begin{figure}[h!]
\centering
\includegraphics[width=0.85\linewidth]{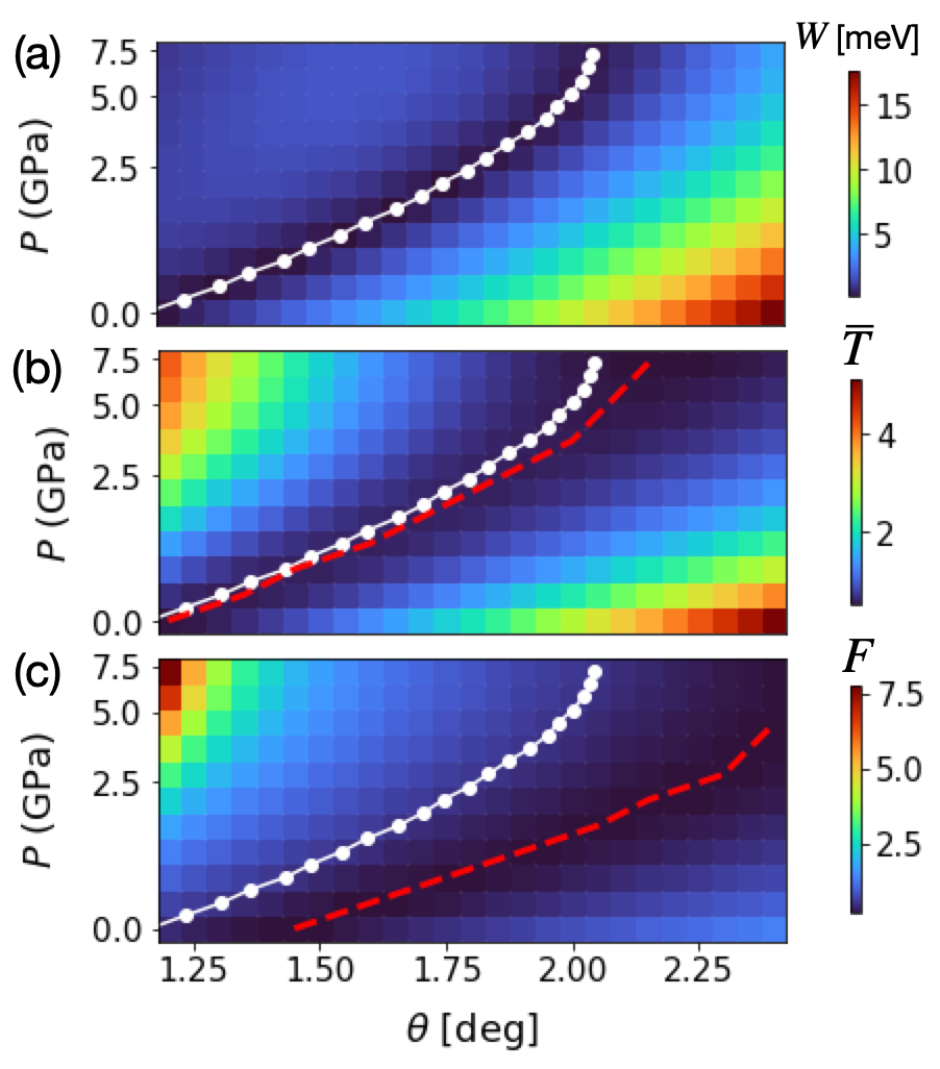}
\caption{FCI single-particle indicators of the topmost moiré valence band of WSe$_2$, plotted as a function of twist angle and pressure: (a) bandwidth $W$, (b) deviation from the trace condition $\overline{T}$ and (c) Berry curvature fluctuations $F$. The white line in each plot corresponds to the {\it magic line} where bandwidth is minimized. Dashed red lines in (b)-(c) trace the minimum value of the quantity being plotted. There is a region of the phase diagram where the three indicators are simultaneously close to zero.} 
\label{fig:Bandwidth_Trace}
\end{figure}

Fig.~\ref{fig:Bandwidth_Trace}(a) shows the bandwidth of the topmost moiré valence band as a function of twist angle and pressure. At vanishing pressure, the bandwidth is a non-monotonic function of twist angle, exhibiting a minimum at a `magic angle' \cite{FengchengTopology, LiangFuMagic}. Fig.~\ref{fig:Bandwidth_Trace}(a) shows that the minimum extends over the full range of pressures considered, resulting in the appearance of a {\it magic line}, indicated in white and new to this work.

%Although a flat Chern band favors a FCI ground state, it does not guarantee it. 
The stability of the FCI phase will also be impacted by the geometric properties of the Bloch wavefunctions, as contained in the quantum geometric tensor
\begin{align}
    Q^{ab}_{\bm k}=\braket{D^a_{\bm k} u_{\bm k}|D^b_{\bm k} u_{\bm k}}=g^{ab}_{\bm k}+\frac{i}{2}\, \epsilon^{ab}\Omega_{\bm k},
    \label{QuantumTensor}
\end{align}
where $D^{a}_{\bm k} \equiv \partial^a_{\bm k} - i A^a_{\bm k}$ is the covariant derivative, $A_{\bm k}^a = -i\langle u_{\bm k}|\partial^a_{\bm k}u_{\bm k}\rangle$ is the Berry connection and $u_{{\bm k}}$ is the periodic part of the Bloch function. The right-hand-side of Eq. \eqref{QuantumTensor} expresses the quantum geometric tensor in terms of the Fubini--Study metric $g_{\bm k}^{ab}$ and the Berry curvature $\Omega_{\bm k}$, which satisfy the inequality $\omega_{{\bm k}, ab}\, g^{ab}_{\bm k}\ge \,|\Omega_{\bm k}|$ \cite{Roy_Ideal}, for each momentum $\bm k$ and unit-determinant matrix $\omega_{{\bm k}, ab}$. A band for which a $\bm k$--independent matrix $\omega_{ab}$ exists that saturates the previous inequality, i.e., $\omega_{ab}\, g^{ab}_{\bm k}=\Omega_{\bm k}$, is said to satisfy the generalized trace condition; a dispersionless band satisfying this condition is called an {\it ideal flat band} \cite{Ideal_Jie}.
%\begin{align}
%    \omega_{ab}\, g^{ab}_{\bm k}=\Omega_{\bm k}.
%    \label{IdealCondition}
%\end{align}
The Bloch functions of an ideal flat band admit a universal analytical form closely related to the LLL wave functions \cite{Ideal_Jie,Jie_originmodelFCI,Ideal_Eslam,Mera23,Valentine23}. As a consequence, FQH--like ground states are stable under short--ranged interactions in ideal flat bands.

Beyond the LLL, models satisfying the ideal band condition include the chiral model of TBG \cite{ChiralTBG,LedwithFCI,HierarchyIdeal_Jie,FamilyIdeal_Ledwith,Nodal_Jie}, Dirac fermions in a non-uniform magnetic field \cite{JunkaiJWLF22}, the Kapit-Mueller model \cite{KapitMuller,Bergholtz_KMGeometry} and periodically strained quadratic band materials \cite{wan2022topological}. 
Engineering realistic systems with bands close to the ideal limit \cite{leeEngineeringIdeal2017,eslam_strain22,wan2023nearly,ghorashi2022topological} is a promising direction in the search for FCIs. We now show that moir\'e TMDs host near-ideal flat bands.

We quantify the deviation of moiré TMD bands from the generalized trace condition by \cite{Moller_Ideal, ParameswaranRoy_Ideal,Roy_Ideal, BauerGeometry2016, simonGeometry2020,Ashvin_FCI1,Ashvin_FCI2,Ideal_Jie}
\begin{align}
    \overline{T}=\int_{\text{BZ}} \frac{d^2{\bm k}}{A_{BZ}} \left( \overline{\omega}_{ab}\,g^{ab}_{\bm k} -\Omega_{\bm k} \right),
    \label{TraceDeviations}
\end{align}
where $A_{BZ}$ is the area of the Brillouin zone and $\overline{\omega}_{\,ab}=\omega_{a}\omega_{b}^*+\omega_{a}^*\omega_{b}$
is obtained from the eigenvector $\omega_a$ of $\overline{Q}^{\,ab}$ with the smallest eigenvalue; $\,\overline{Q}^{\,ab}$ indicates the Brillouin-zone averaged quantum geometric tensor. The behavior of $\overline{T}$ with pressure and twist angle is shown in Fig. \ref{fig:Bandwidth_Trace}(b), which reveals that the line of minimum $\overline{T}$ nearly coincides with the magic line defined by minimum bandwidth.

Finally, the LLL also exhibits $\bm k$-independent Berry curvature. Fig. \ref{fig:Bandwidth_Trace}(c) shows the Berry curvature fluctuations 
\begin{align}
    F=\left[ \int_{BZ}\frac{d^2{\bm k}}{A_{BZ}} \left( \frac{\Omega_{\bm k}}{2\pi}-C\right)^2\right]^{1/2},
\end{align}
where $C$ is the Chern number, as a function of twist angle and pressure.  
Notably, while the lines of minimum bandwidth and trace deviations almost coincide, the line of minimum Berry curvature fluctuations is distinct. In an ideal band, such as those of chiral TBG, $W$, $\,\overline{T}$, and $F$ all vanish simultaneously.

{\em Pressure--twist angle phase diagram --- } Motivated by the existence of the region in Fig. \ref{fig:Bandwidth_Trace} where $W$, $\overline{T}$ and $F$ are simultaneously small, we proceed to study many-body ground states in the pressure-{\it vs}-twist angle phase space at fractional band filling. Our approach is to write the fully interacting Hamiltonian in momentum space, where the single-particle term comes from the band energies of the continuum model and interactions are projected onto the topmost moiré band. This projection reduces the Hilbert space, allowing for exact diagonalization to obtain the ground state and excited states. Labelling the single-particle energies and eigenstates from the topmost moir\'e band of Eq. \eqref{BMHamiltonian} as $\epsilon_{\bf k}$ and $\ket{u_{\bf k}}$, respectively, the interacting Hamiltonian reads
%
%\begin{align}
%\label{ManyBodyHamiltonian}
%    H=\sum_{{\bm k}}&\epsilon_{{\bm k}}c^{\dagger}_{{\bm k}\sigma}c_{{\bm k}\sigma}\nonumber \\
%    &+\frac{1}{2}\sum_{\substack{{\bm k},{\bm k'}{\bm q}\\ \sigma,\sigma^{\prime}}}V_{{\bm k},{\bm k'},{\bm q}}^{\sigma,\sigma^{\prime}}c^{\dagger}_{{\bm k},\sigma}c^{\dagger}_{{\bm k'},\sigma^{\prime}} c_{{\bm k'-{\bm q}},\sigma^{\prime}}c_{{\bm k}+{\bm q},\sigma}.
%\end{align}
%
\begin{align}
\label{ManyBodyHamiltonian}
    H \! = \! \sum_{{\bm k}}&\epsilon_{{\bm k}}c^{\dagger}_{{\bm k}\sigma}c_{{\bm k}\sigma}+\!\frac{1}{2}\! \sum_{\substack{{\bm k},{\bm k'}{\bm q}\\ \sigma,\sigma^{\prime}}}V_{{\bm k},{\bm k'},{\bm q}}^{\sigma,\sigma^{\prime}}c^{\dagger}_{{\bm k},\sigma}c^{\dagger}_{{\bm k'},\sigma^{\prime}} c_{{\bm k'-{\bm q}},\sigma^{\prime}}c_{{\bm k}+{\bm q},\sigma}.
\end{align}
Here $\sigma/\sigma'$ are valley indices. The gate-screened Coulomb interaction elements projected to the topmost band are 
\begin{align}
    V_{{\bm k},{\bm k'},{\bf q}}^{\sigma,\sigma^{\prime}}=\frac{1}{A} \sum_{{\bm G}}\Lambda_{{\bm k},\sigma}^{{\bf q}+{\bm G}}\Lambda_{{\bm k'},\sigma'}^{-{\bf q}-{\bm G}}\,\frac{2\pi e^2}{\varepsilon\, \tilde{q}}\tanh \left( \tilde{q}\,d\right),
\end{align}
where $\tilde{q}=|{\bm q}+{\bm G}|$ is the momentum transfer, $A$ is the system area, $\Lambda_{{\bm k},\sigma}^{{\bm q}+{\bm G}}=\braket{u_{{\bm k},\sigma}|u_{{\bm k}+{\bm q}+{\bm G},\sigma}}$ are the form factors, $\varepsilon$ is the dielectric constant and $d$ denotes the distance from the sample to metallic gates. For now, we fix $d=10$ nm and $\varepsilon=30$ to remain well within the validity of our band projection.
In the supplemental material \cite{Supplemental}, we show results for $\varepsilon=10$, closer to the experimental value; the phase diagram displays the same qualitative features. 
 
\begin{figure}[h!]
\centering
\includegraphics[width=\linewidth]{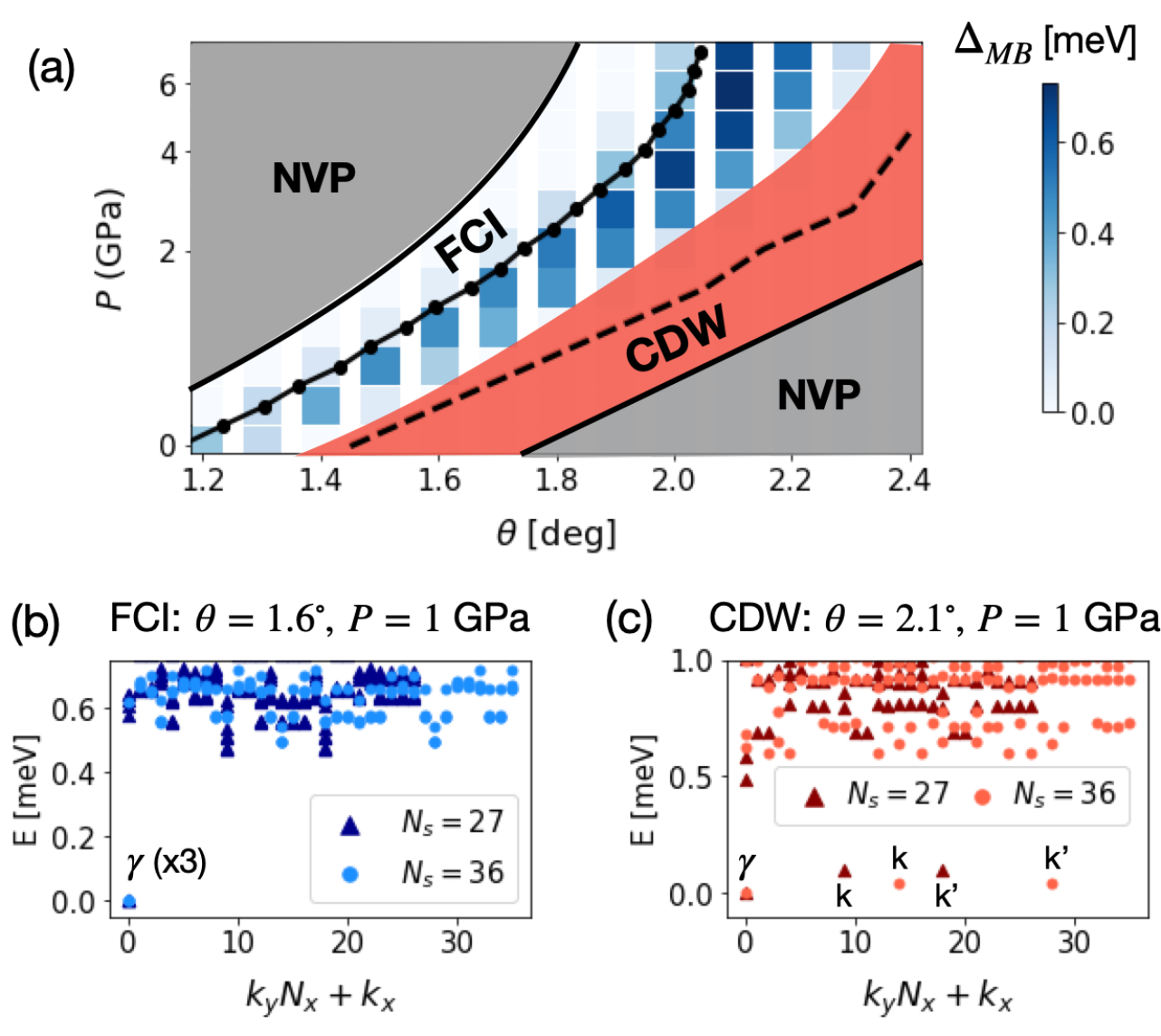}
\caption{(a) Phase diagram of twisted bilayer WSe$_2$ at filling $\nu=1/3$ as a function of twist angle and pressure. Gray indicates that the ground state is not valley-polarized (NVP). Blue and red indicate valley-polarization and correspond to an FCI and a CDW, respectively. Blue color scale indicates the many-body gap $\Delta_{MB}$ in the FCI phase for $N_s=27$. The solid line with dots is the magic line and the dashed line indicates the minimum of $F$. (b) Many-body spectrum for a representative FCI as a function of linearized many-body momentum, displaying a threefold-degenerate ground state at ${\bm \gamma}$. (c) Many-body spectrum for a CDW state, exhibiting three nearly-degenerate ground states with momenta ${\bm \gamma}$, ${\bm k}$, ${\bf k}^{\prime}$.} 
\label{fig:PhaseDiagram_FCI}
\end{figure}

We diagonalize the Hamiltonian in Eq. \eqref{ManyBodyHamiltonian} at band filling factor $\nu=N/N_s=1/3$, where $N$ is the number of holes and $N_s$ the number of moiré unit cells, for different finite system sizes $N_s$. The resulting phase diagram as a function of twist angle and pressure is shown in Fig. \ref{fig:PhaseDiagram_FCI}(a). 
The colored area between the solid black lines indicates the region where Coulomb interactions induce a fully valley-polarized ground state. 

Within the valley-polarized regime, the blue region indicates a ground state in the same universality class as the Laughlin state, i.e., an FCI, which extends over approximately half a degree. The FCI is identified by the many-body spectrum in Fig. \ref{fig:PhaseDiagram_FCI}(b), showing a three-fold degenerate ground state with a clear gap to excited states. 
Upon flux insertion, the three degenerate ground states evolve into each other and remain separated from all excited states \cite{Supplemental}.

Fig.~\ref{fig:PhaseDiagram_FCI} reveals two important insights:  First, applied pressure both extends the range of angles over which the FCI phase can be realized and increases the many-body gap.
Second, the FCI is centered around the {\it magic line} where $W$ and $\overline{T}$ are minimized, while the CDW is centered around the line of minimum Berry curvature fluctuations. This highlights the limits of Berry curvature as an FCI indicator, since minimizing $F$ does not always increase FCI stability \cite{Bergholtz_KMGeometry}.
 
Adjacent to the FCI phase and within the valley-polarized region is a competing CDW ground state, which dominates when bandwidth and trace deviations become large enough. The many-body spectrum of the CDW phase, shown in Fig. \ref{fig:PhaseDiagram_FCI}(c), has a characteristic three-fold ground state degeneracy consisting of states with many-body momentum ${\bm \gamma}$, ${\bf k}$ and ${\bf k}^{\prime}$.

\begin{figure}[h!]
\centering
\includegraphics[width=\linewidth]{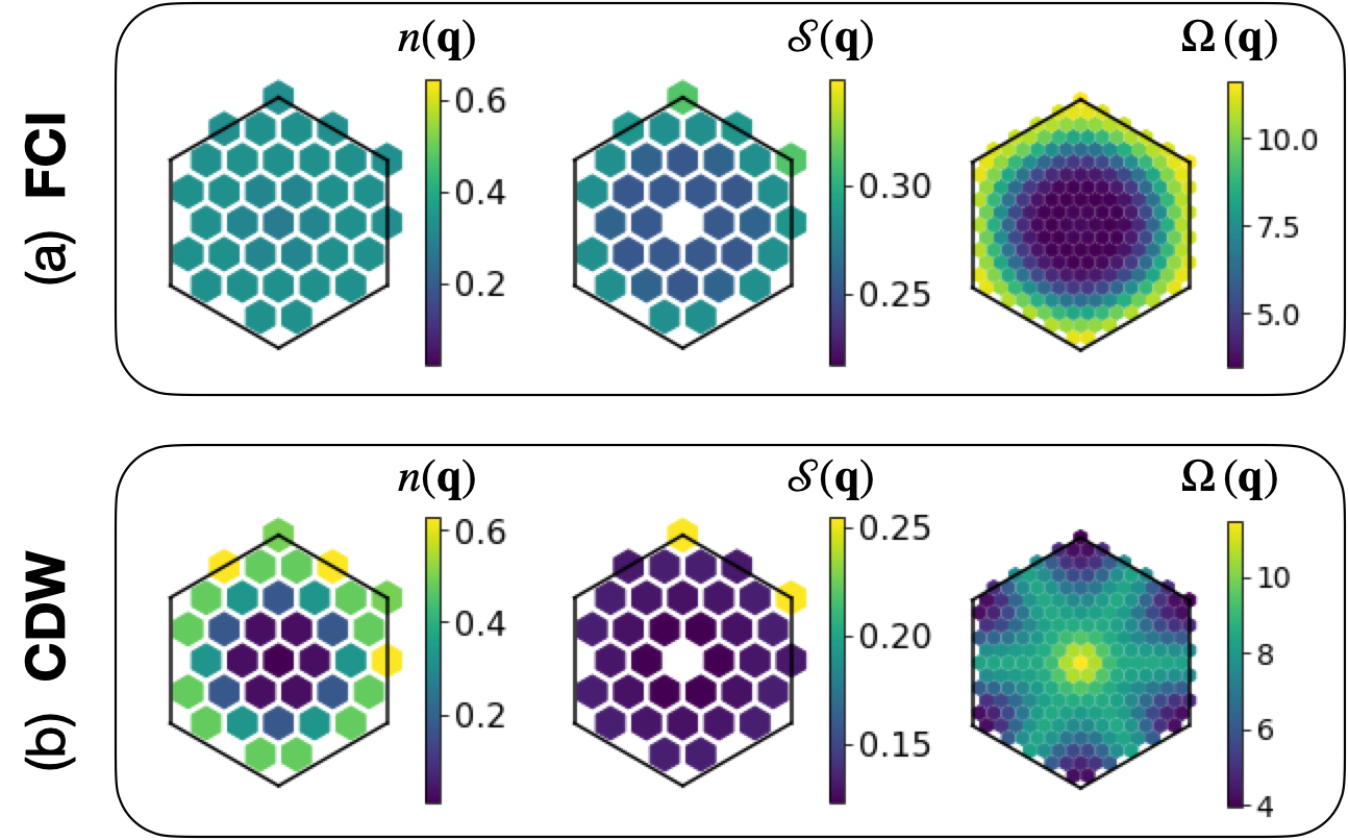}
\caption{Ground state occupation $n({\bm q})$, static structure factor $\mathcal{S}({\bm q})$ and Berry curvature $\Omega({\bm q})$ for (a) the FCI ($\theta=1.6^\circ$, $P=1$ GPa) and (b) the CDW ($\theta=2.1^\circ$, $P=1$ GPa). Results for occupations and structure factors were calculated using $N_s=36$. The peak in the structure factor at $\bm \gamma$ corresponding to the valley-polarization has been omitted to highlight the structure at other momenta.} 
\label{fig:FCI_CDW}
\end{figure}

To further characterize the FCI and CDW phases, Fig. \ref{fig:FCI_CDW} shows the occupation $n({\bm q})$, static structure factor $\mathcal{S}({\bm q})$ and Berry curvature $\Omega({\bm q})$ for representative points within each phase. In the FCI phase, the charge occupation and structure factor are nearly constant, characteristic of a Laughlin-like state. In contrast, in the CDW phase, the occupation increases towards the edge of the Brillouin zone, while the structure factor shows peaks at the ${\bf k}/{\bf k}^{\prime}$ points, indicating moir\'e translation symmetry breaking to a $\sqrt{3}\times\sqrt{3}$ state. The Berry curvature in the CDW phase is peaked around ${\bm \gamma}$, where the ground state occupation number vanishes. Thus, the charge carriers do not strongly feel the effective magnetic field, consistent with its trivial topology.
In contrast, the ground state occupation in the FCI phase is more uniformly distributed, so that charge carriers feel the Berry curvature throughout the Brillouin zone, resulting in topological order. A similar trend was found in TBG \cite{Lauchli_FCI,Ashvin_FCI1,QuantumMetric_Bergholtz}.
However, one difference is that in TBG the Berry curvature is always peaked at ${\bm \gamma}$, while for TMD homobilayers the peak in Berry curvature moves as the model parameters vary, as seen in Fig. \ref{fig:FCI_CDW}. 

%It has been suggested that quantum geometry plays a prominent role in dictating the ground state properties in interacting topological bands, possibly triggering topological phase transitions \cite{Repellin_FCI_TBG,QuantumMetric_Bergholtz}. 
{\em Discussion --- } 
%The growing field of topological moiré materials provides a fertile scenario to experimentally study topological order without a strong magnetic field. From a theoretical perspective, there is much room for progress in this area, with possibilities that range from identifying qualitatively new phases of matter beyond the fractional quantum Hall paradigm, to accomplishing a deeper understanding of how discrete translation symmetry affects topological order.
%
We have proposed semiconductor moiré materials as systems with rich quantum geometry that can be tuned experimentally via applied pressure to stabilize topologically-ordered phases. Specifically, in twisted WSe$_2$ the `magic angle' where the bandwidth is minimized at zero pressure turns into a {\it magic line} with similarly small bandwidth at finite pressures. The {\it magic line} extends the range of angles over which the FCI phase is stable.
Further, the quantum geometry of the band in the region around the {\it magic line} is nearly ideal for realizing an FCI. 

We provide numerical results that support an FCI ground state, extending previous studies \cite{KaiSunFCI, LiangFuFCI} to finite pressure. Further, the FCI phase realized at finite pressure has a larger many-body gap than that at zero pressure. Experimentally, we predict that at small twist angles, if an incompressible CDW phase is measured in a moiré TMD homobilayer, applying pressure could drive a transition into an FCI phase. The pressure needed is within experimental reach. Specifically, applied pressures of up to $P \sim 2$ GPa, or equivalently 5$\%$ compression, have been realized in TBG \cite{CoryPressure1,CoryPressure2}. 
In TMDs, pressures up to $P \sim 5$ GPa have been realized at room temperature, and $P \sim 1.4$ GPa \cite{Pressure_TMD_Berkeley} at cryogenic temperatures. 
Although Chern bands have not been conclusively observed in twisted WSe$_2$ \cite{ColumbiaWSe}, experiments on heterobilayers \cite{Cornell_QAH1,Cornell_QAH2,Cornell_QAH3} have observed a Chern insulator at half-filling, though in a regime of large bandwidth unfavorable for realizing the FCI ground state.

As in TBG, the intertwined effects of band dispersion, quantum geometry, and long-range electronic interactions in our model combine to ultimately determine the ground state properties. Despite these complex factors, in TBG the chiral model has provided powerful analytical insight \cite{ChiralTBG, Ideal_Jie}. 
The proximity of our model to the ideal condition motivates a future search for a chiral model of moir\'e TMDs in a suitable limit.
More generally, the question of how perturbations from the ideal limit affect an FCI ground state and its excitations is an open one. 
The near-ideal Chern band we have described in moir\'e TMDs suggests that approximating the ground state wavefunctions by modified Landau level wavefunctions is a reasonable approximation \cite{Ideal_Jie,Ashvin_FCI1} upon which to build a perturbative study of non-ideal Chern bands.
\newline

{\em Acknowledgments ---} We thank Allan H. MacDonald and Andrew J. Millis for insightful discussions. We acknowledge financial support and computing resources from the Flatiron Institute, a division of Simons Foundation. NMD was supported by the CCQ Pre-Doctoral program. GRS and EK acknowledge funding from the Army Research Office under award W911NF-21-2-0147 and from the Simons Foundation award No. 896626. ZZ acknowledges support from a Stanford Science Fellowship. CR acknowledges support by the Flatiron Institute IDEA Scholar Program. JC acknowledges the Alfred P. Sloan Foundation through a Sloan Research Fellowship. This work was performed in part at Aspen Center for Physics, which is supported by National Science Foundation Grant No. PHY-2210452, and in part at the Kavli Institute for Theoretical Physics, supported by the National Science Foundation under Grant No. NSF PHY-1748958. 

\bibliography{refs}

\newpage
\appendix
\onecolumngrid
\section*{Supplemental material for ``Pressure--enhanced fractional Chern insulators in moir\'e transition metal dichalcogenides along a magic line" }
\twocolumngrid
\section*{\textit{Ab initio} continuum model parameters as a function of pressure}
We used density functional theory (DFT) calculations to investigate the electronic and structural properties of WSe$_2$ under external pressure. Specifically, we utilized the VASP DFT software \cite{VASP96} and projector-augmented wave (PAW) pseudopotentials \cite{PAW,PAW_kresse} to describe the electron-ion interactions. To accurately capture the van der Waals interactions in the 2D material, we employed the r$^2$SCAN-D3(BJ) meta-GGA exchange-correlation functional, shown to provide accurate results for systems with weakly interacting layers \cite{r2SCAN,r2SCAN-D4}.
To ensure the accuracy of our DFT calculations, we used an energy cutoff of 292 eV (1.3 $\times$ the maximum energy cutoff specified by pseudopotential, as defined in the Materials Project accuracy standard \cite{Jain2011}) and a \textbf{\textit{k}}-point density of 21$\times$21$\times$1 $\Gamma$-centered Monkhorst-Pack meshes to sample the Brillouin zone. We set the energy threshold for self-consistent field (SCF) convergence to 10$^{-5}$ eV and the relaxation threshold for ionic relaxation of Hellmann-Feynman forces smaller than 10$^{-3}$ eV/Å. To avoid spurious interactions with periodic images, we included a vacuum layer of 15 Å in the direction perpendicular to the 2D material.

We first obtained the relaxed lattice parameter of bilayer WSe$_2$, 3.279 Å, by performing atomic relaxations without spin-orbit coupling (SOC) until the forces on the atoms were below the convergence threshold. The relaxed vertical separation between the metal atom planes in each layer is 7.06 Å and 6.51 Å for the AA and MX stackings, respectively. 
Next, we evaluated the pressure effects on the WSe$_2$ bilayer by fixing the z-coordinate of the lowermost atomic layer and compressing the topmost atomic layer vertical distance by 4\% and 10\%, relaxing atomic positions with this displacement constrain, and then measured the resulting pressure in GPa, for both high-symmetry AA and MX/XM bilayer stackings. 
We then obtained a converged self-consistent charge density with spin-orbit coupling (SOC) and calculated the band structure of WSe$_2$ with SOC included. 
Finally, with the eigen-energies obtained from the band structure calculations aligned relative to the vacuum level, we fit the continuum model parameters $(V_m, \psi, \omega)$ using our \textit{ab initio} results \cite{FengchengTopology, Mattia_Gamma_TMDs}.
For twisted semiconductors, the moiré bands closest to the band gap can be derived from the untwisted parabolic band extrema (or valleys) in the original Brillouin zone \cite{twist_JanusTMDs}. In twisted WSe$_2$, the valley close to charge neutrality (and hence explored in experiments) is $K$ \cite{FengchengTopology, Mattia_Gamma_TMDs}.

\begin{figure}[ht!]
\centering
\textbf{(a)} AA stacking \\
\includegraphics[width=0.85\linewidth]{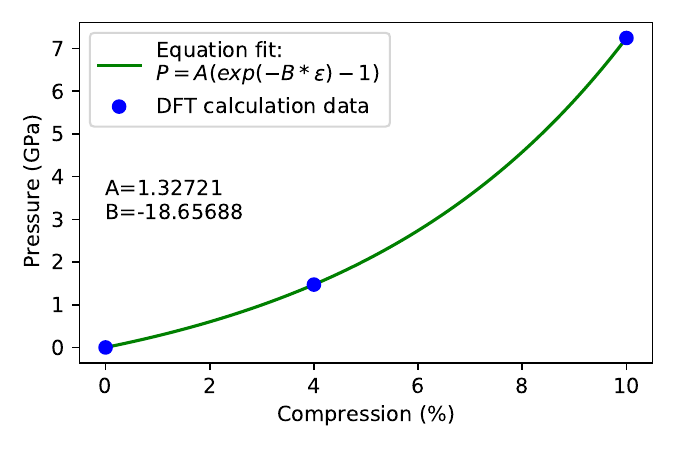}\\
\textbf{(b)} MX/XM stacking \\
\includegraphics[width=0.85\linewidth]{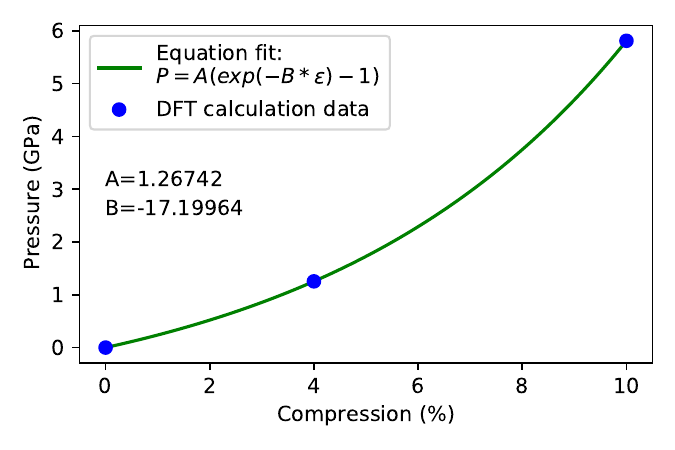}
\caption{Fitting of the calculated vertical external pressure (green line) as a function of the interlayer distance between WSe$_2$ layers (blue points) for \textbf{(a)} AA and \textbf{(b)} MX/XM bilayer stackings.}
\label{fig:Pressure_vs_compression}
\end{figure}

The pressure obtained from the DFT calculations can be fitted by the expression \cite{PressureTBG}
\begin{equation}
    P=A(e^{-B \epsilon}-1) \ ,
\end{equation}
where $\epsilon$ is the compression, obtaining for bilayer WSe$_2$ the values of A = 1.327 GPa and B = -18.657 for the AA stacking, and A = 1.267 GPa and B = -17.200 for the MX/XM stacking, both displayed in Fig. \ref{fig:Pressure_vs_compression}.
Regarding the pressure effect on the DFT-extracted continuum model parameters $(V_m, \psi, \omega)$ shown in Fig. 1, they can be described by a quadratic fit \cite{PressureTBG}: 
\begin{equation}
  y=c_0+c_1 \epsilon+c_2 \epsilon^2 \ .    
  \label{eq:pressure_params}
\end{equation}
The fitted values are given in Table \ref{table:pressure_continuum_params}.
\begin{table}[h!]
\caption{Fitting of the calculated continuum model parameters as a function of pressure, as given by Equation \ref{eq:pressure_params} and shown in Figure \ref{fig:BandstructureParameters}.}
\begin{ruledtabular}
\begin{tabular}{lddd}
 & c_0 & c_1 & c_2 \\ \hline
$V_m$ & 6.359 & 1.3982 & 0.0456 \\
$\psi$ & 115.729 & -0.6314 & -0.0829 \\
$\omega$ & 8.897 & 1.3860 & 0.0811
\label{table:pressure_continuum_params}
\end{tabular}
\end{ruledtabular}
\end{table}
Compression of the bilayer has a strong effect on the interlayer coupling and moiré potential, since the electronic coupling between the layers increases as the layers are forced closer together. This trend can be expected for other materials whenever the involved band edges (valence or conduction) are composed of out-of-plane orbitals.

%What is the argument to neglect strain in these calculations? For instance, substrate.
%-- Show the Berry curvature trends for the three different functionals at zero pressure as a function of twist angle? -- Depending on the details of the calculation the phase diagram could change quantitatively but the trend should be the same.

Finally, we perform a careful evaluation of the sensitivity of the continuum model parameters to the DFT calculations parameters, most importantly the exchange-correlation functional used and the reference value for the energies, aligning to the vacuum level or not. The results are presented in Table \ref{table:dft_functionals}. In preliminary calculations, we also check that pseudopotentials including semicore states do not lead to significant changes in results.
The vacuum level alignment changes the relative energies between AA and MX states, thus directly increasing $\psi$ and also moderately increasing $V_m$. Different functionals change the magnitude of the model parameters in proportion to their tendency to increase or decrease interaction/overlap strengths. For instance, LDA and PBE are widely known DFT functionals, known to over- and under-bind systems, respectively.
\begin{table}[h]
\caption{Comparison of the calculated continuum model parameters $(V_m, \psi, \omega)$ as a function of the DFT exchange-correlation functional used in calculations. \label{table:dft_functionals}}
\begin{ruledtabular}
\begin{tabular}{lddd}
DFT functional & V_m & \omega & \psi \\ \hline
LDA & 8.754 & 9.462 & 114.532 \\
$\hookrightarrow$ without vacuum alignment & 7.964 & 9.462 & 90.000 \\
r$^2$SCAN-D3(BJ) & 6.359 & 8.897 & 115.729  \\ 
$\hookrightarrow$ without vacuum alignment & 5.729 & 8.897 & 90.000 \\
SCAN+rVV10 & 5.456 & 8.031 & 116.438  \\ 
$\hookrightarrow$ without vacuum alignment & 4.885 & 8.031 & 90.000 \\ \hline
Ref. \cite{FengchengTopology} LDA\footnote{The eigenvalues are not aligned from a calculated vacuum level, therefore $\psi \approx \pi/2$.} & 8.900 & 9.700 & 91.000  \\
$\hookrightarrow$ + vacuum level alignment\footnote{Artificial AA/MX vacuum (mis)alignment added.} & 9.781 & 9.700 & 114.526  \\
Ref. \cite{LiangFuMagic} SCAN+rVV10 & 8.670 & 18.000 & 125.113 \\
$\hookrightarrow$ + swapping AA/MX\footnote{From the supporting information of ref. \cite{LiangFuMagic}, the unusual values reported in comparison to previous results of ref. \cite{FengchengTopology} suggest that the eigenvalues for AA and MX are swapped in fitting the model parameters.}  & 11.527 & 12.284 & 115.634  \\
\end{tabular}
\end{ruledtabular}
\end{table}
Another subtle effect of the functionals that change the models results are the absolute energies of the AA and MX states in relation to vacuum level. For WSe$_2$, the energy difference $\Delta_{AA/MX}$ is on the order of a few dozens of meV (MX states are higher in energy). Our results show that for LDA $\Delta_{AA/MX} = 36.2$ meV, for r$^2$SCAN-D3(BJ) $\Delta_{AA/MX}=26.71$ meV, and for SCAN+rVV10 $\Delta_{AA/MX}=21.57$ meV. Additionally, $\Delta_{AA/MX}$ increases with pressure.
The implication from these validations is that depending on the details of the calculation, the magnitude of the parameters may change, and accordingly the phase diagram could also change quantitatively, but the trends should be the same.
%\gs{I think adding the figures of Berry curvature trends for the three different functionals at zero pressure as a function of twist angle would be a good addition.}

\section*{Gap to remote bands and flat band projection}
\begin{figure}[t!]
\centering
\includegraphics[width=\linewidth]{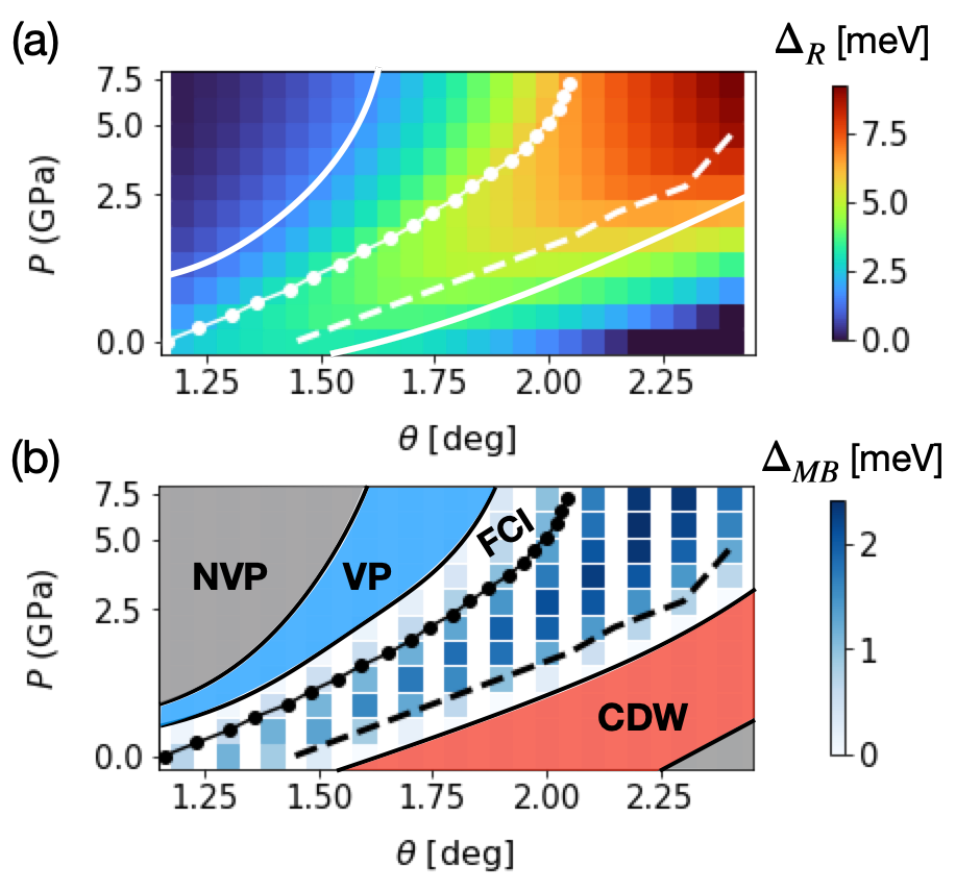}
\caption{(a) Gap between the first and second moiré valence bands, $\Delta_R$, as a function of twist angle and pressure. The solid white lines enclose the region where the bandwidth is smaller than $\Delta_R$. (b) Phase diagram of twisted bilayer WSe$_2$ at filling $\nu=1/3$ and for $\varepsilon=10$, as a function of twist angle and pressure. The magic line is indicated as the black line with dots, while the minimum of $F$ is the black dashed line.}
\label{fig:Gap_Remoteband}
\end{figure}

Given the exponential growth of the Hilbert space with system size, our approximation of projecting interactions to the topmost band is required for application our exact diagonalization (ED) method to sensible system sizes. This approximation is better justified when the gap between the first and second moiré valence bands, $\Delta_R$, is larger than the other energy scales of the problem, namely the bandwidth $W$ and the interaction energy scale, given by $U_M=e^2/\varepsilon a_M$. In Fig. \ref{fig:Gap_Remoteband}(a) we show $\Delta_R$ for WSe$_2$ as a function of twist angle and pressure. In the middle region, delimited by solid white lines, the gap $\Delta_R$ is larger than the bandwidth and larger or equal than $U_M$ for the parameter $\varepsilon=30$ used for the phase diagram presented in the main text. This justifies the band projection in the regime of interest, where we find valley-polarized states, as seen in Fig. \ref{fig:PhaseDiagram_FCI}(a) of the main text. 
It should also be noted that $\Delta_R$ increases with applied pressure, which may improve the conditions for stabilizing the FCI at a given interaction strength.

Experimental values for samples encapsulated between hBN are in the range $\varepsilon \sim 5-10$. As commented above, in the main text, we have considered a larger value of $\varepsilon = 30$ to ensure the validity of band projection. For completeness, Fig. \ref{fig:Gap_Remoteband}(b) shows the phase diagram obtained with the dielectric constant $\varepsilon=10$. For this interaction strength the region of valley-polarization increases. In particular the FCI phase now contains the magic line and the line of minimum Berry curvature fluctuations. To the left of the FCI we also find a region of valley-polarization (labelled VP and colored in solid blue in Fig. \ref{fig:Gap_Remoteband}(b)), whose
ground state is non-degenerate with many-body momentum $\bm \gamma$, thus not corresponding to either the FCI or CDW phases. The many-body gap at $\varepsilon = 10$ is significantly larger than at $\varepsilon=30$, as can be seen by the color scale. This suggests that stronger Coulomb interactions favor topological order, while the translation symmetry breaking is due to bandwidth being the dominant energy scale. 
%The FCI could be favored over the CDW if one could engineer a system with a sufficiently flat band.

The role of screening is also important in stabilizing the FCI state. We used a screened interaction, since strong long-range interactions could promote level mixing, destroying the FCI state \cite{LevelMixing}.
It has also been argued that for short-ranged interactions, topological order can be stable even when interaction strength exceeds the gap to remote bands \cite{Kourtis}, which would further increase the validity region of our projection approach. 
The moiré period is given by $a_M\approx a_0/\theta$, where $a_0=0.3279$ nm is the lattice constant of monolayer WSe$_2$. For the range of angles considered, the moiré period is in the range $7.8-15.7$ nm, hence in all our calculations we placed the metallic gates at $10$ nm, which is of the same order as the moiré length in order to screen long-range Coulomb interactions. 

\section*{Further details on exact diagonalization}
To determine the region of full valley-polarization due to Coulomb interactions, we performed exact diagonalization calculations on $N_s=24$ moiré unit cells. For this system size, $\nu=1/3$ filling corresponds to populating 8 holes into the band and we can calculate all possible configurations --$(N_K,N_{K^{\prime}})$-- with $N_K$ holes on valley $K$ and $N_{K^{\prime}}$ holes on valley $K^{\prime}$. A system size for which all $(N_K,N_{K^{\prime}})$--sectors can be accessed is important when checking valley-polarization, because it may happen that the ground state switches directly from the fully-polarized sector to the one with an equal number of carriers in each valley, which we refer to as the time-reversal symmetric (TRS) sector. In such case, calculating only the valley-flips would not give the correct phase boundary. We obtain that valley-polarization is lost both to the left and the right of the phase diagram for $\varepsilon=30$ and $\varepsilon=10$. These regions are colored in gray and labelled NVP. In the NVP region to the right of the CDW phase, we attribute the loss of polarization to the increase in bandwidth. This leads to the kinetic energy dominating over Coulomb exchange, which disfavors valley-polarization. 

The phase to the left of the FCI occupies a significant part of the phase diagram and is more complicated to interpret. We find that in the NVP phase on the left of the phase diagrams, besides the ground state not being valley-polarized at $N_s=24$, the spectrum in the polarized sector does not resemble that of an FCI phase. We note that the regime where this phase appears is where our single band projection is less well-justified due to the close proximity of the second moiré valence band; see Fig. \ref{fig:Gap_Remoteband}(a). Nevertheless, once the projection is done the band is agnostic to remote bands, so the disappearance of the FCI could be attributed to non-ideal band geometry. Because in this regime the two topmost moiré bands have opposite Chern numbers, a treatment that accurately addresses the ground state would involve including two bands. This approach was taken in \cite{LiangFuFCI} by first mapping these two bands to a real space model on a honeycomb lattice and then adding interactions. In that study a similar trend was obtained: The FCI state is stable in a small region around the magic angle and then it is lost due to valley de-polarization.

\begin{figure}[t!]
\centering
\includegraphics[width=0.95\linewidth]{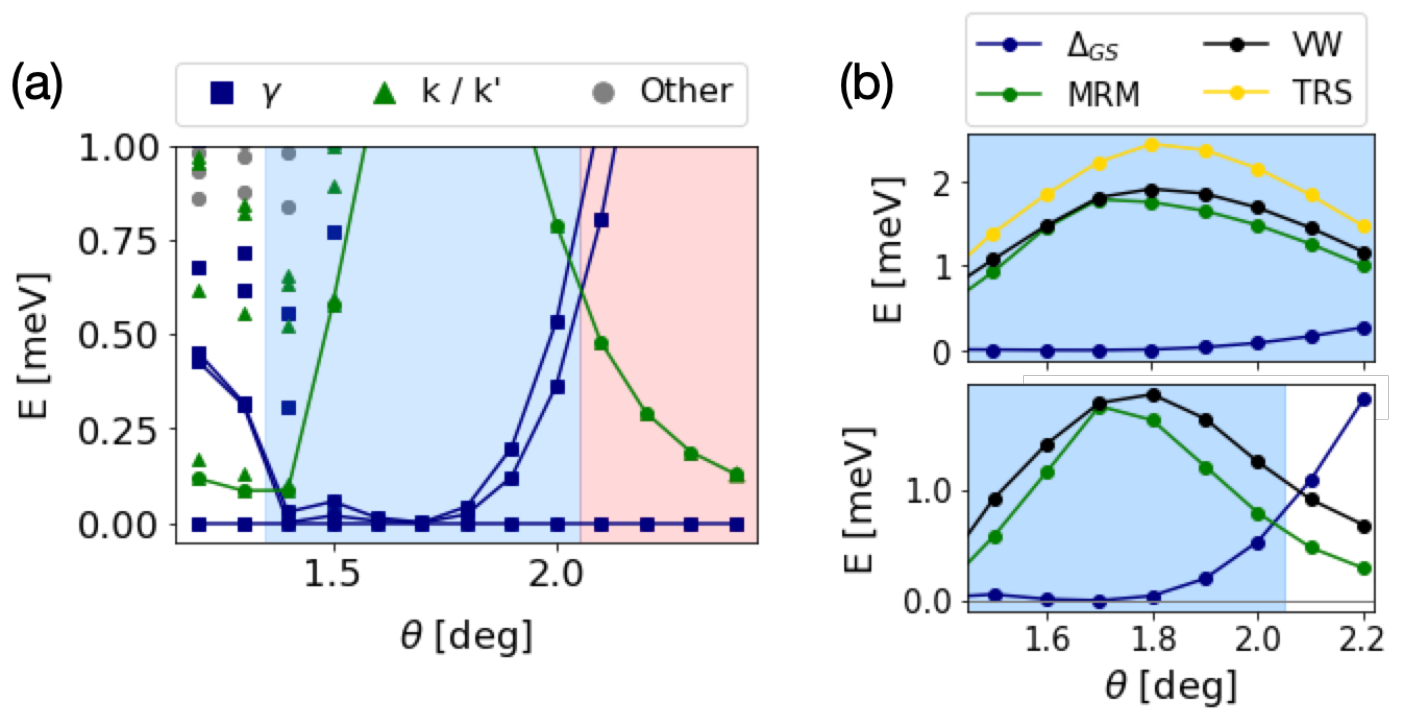}
\caption{(a) Exact diagonalization spectrum for $N_s=27$ as a function of twist angle for $P=1$ GPa (3$\%$ compression). The blue region corresponds to the FCI, with a three-fold degenerate ground state with many-body momentum ${\bm \gamma}$ and the red region to the CDW, with a three-fold quasi-degenerate ground state with many-body momenta ${\bm \gamma}$, ${\bf k}$ and ${\bf k'}$. In the white region topological order is lost. (b) Low-energy excitations that destabilize the FCI phase for $N_s=24$ (top) and (b) $N_s=27$ (bottom). These include the magneto-roton mode, valley-waves and a time-reversal-symmetric excitation (only shown for $N_s=24$).}
\label{fig:Excitations}
\end{figure}

Evolution of the low-energy many-body energies for fixed $P=1$ GPa and as a function of twist angle is shown in Fig. \ref{fig:Excitations}(a), for $N_s=27$ and $\varepsilon=10$. The states belonging to the valley-polarized sector and with many-body momentum ${\bm \gamma}$ and  ${\bf k}$/${\bf k'}$ correspond to the blue and green markers respectively, while the gray markers correspond to all other momenta and also to the sector of valley-waves. The evolution of the three FCI ground states is shown as blue lines and the evolution for the two CDW states at ${\bf k}$/${\bf k'}$ is shown as green lines. Level-crossings indicate different transitions, first from a valley-polarized phase (white, not an FCI) to the FCI (blue) and then to the CDW (red).  

We also study the different low-energy excitations as a function of twist angle in the FCI phase for systems with $N_s=24$ and $N_s=27$ moiré unit cells. These results are shown in Fig. \ref{fig:Excitations}(b). We focus on the neutral excitation or magneto-roton mode, the valley-waves and the lowest excitation in the TRS sector. For both system sizes we see that the first excitation to overcome the ground state and close the FCI gap is the neutral excitation, consistent with the CDW phase taking place. The stability region of the FCI is colored in blue for both sizes in Fig.  \ref{fig:Excitations}(b). For $N_s=24$, due to the absence of the ${\bf k}$-points in the Brillouin zone discretization, the phase extends further. For $N_s=27$ we do not have access to the TRS excitations. 
%It could be that pressure improving the FCI conditions is generic
%Estimate charge gaps? 
%Time reversal symm.  $\mathcal{H}^{K^{\prime}}({\bm k})=\left[ \mathcal{H}^{K}(-{\bm k})\right]^{*}$. 

\begin{figure}[t!]
\centering
\includegraphics[width=\linewidth]{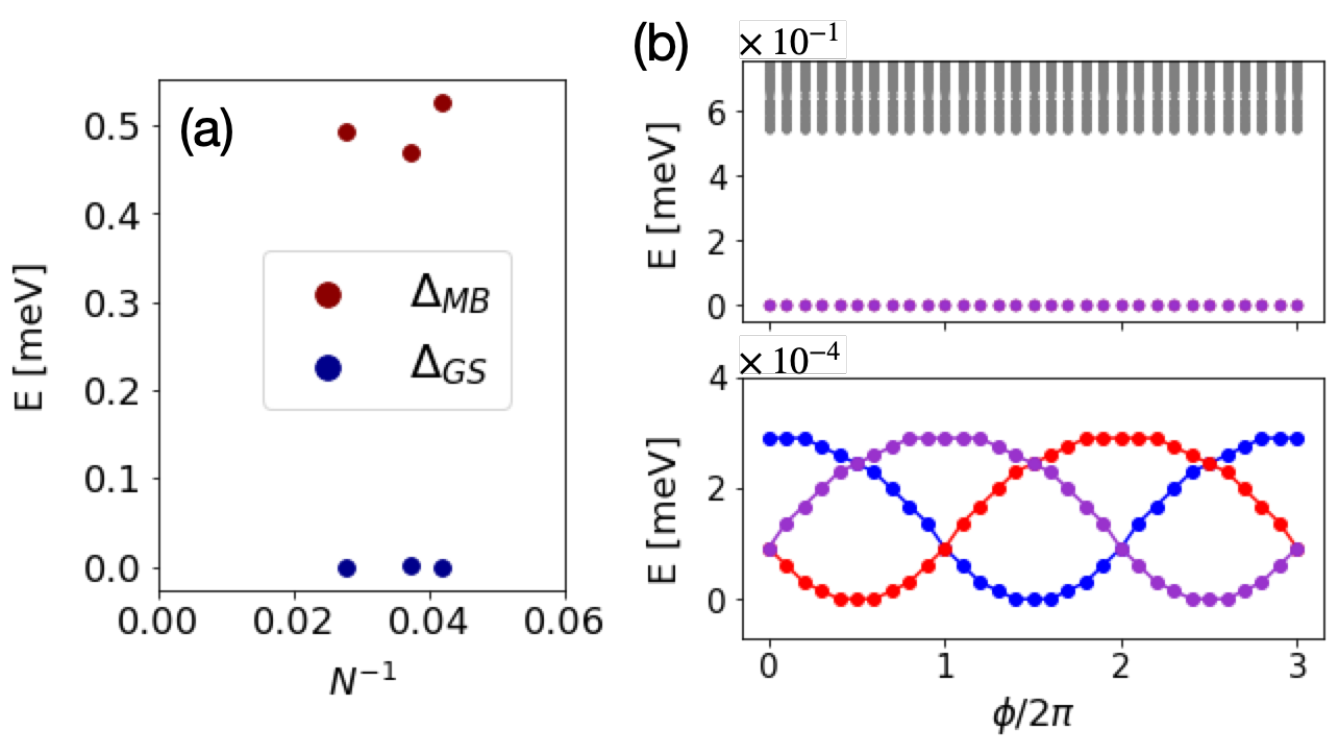}
\caption{Fractional Chern insulator for $\theta=1.6^{\circ}$ and $P=1$ GPa. (a) Scaling of the many-body gap $\Delta_{MB}$ and the energy difference between the three ground states, $\Delta_{GS}$, as a function of system size. (b) Adiabatic flux insertion along ${\bm T}_2$ direction, showing that the three ground states remain isolated from excited states (top) and that they evolve into each other (bottom).}
\label{fig:GapExtrapolation_SpectralFlow}
\end{figure}

To provide further evidence for the topologically-ordered phase, in Fig. \ref{fig:GapExtrapolation_SpectralFlow}(a) we show how the many-body gap $\Delta_{MB}$ and the gap between the three degenerate ground states of the FCI phase $\Delta_{GS}$ scale with system size. The former gap should extrapolate to a finite value in the thermodynamic limit, while the latter should extrapolate to zero. Oscillations in the behavior of the many-body gap are due to performing the ED calculation in the system of size $N_s=24$, that does not contain the corners of the Brillouin zone and hence overestimates this gap. We also plot the evolution of the FCI ground state under adiabatic flux insertion in Fig. \ref{fig:GapExtrapolation_SpectralFlow}(b). It can be seen that the three nearly-degenerate ground states evolve into each other and remain separated from all other excited states upon flux insertion, as expected for a fractional quantum Hall-like phase. We note also that the exact ground state degeneracy of the CDW should be recovered in the thermodynamic limit. 
%How does this relate with the phase called valley-polarized in\cite{KaiSunFCI}.?
%\red{We can also show results for structure factors for system size $N_s$=27. This indicates size-independence of the trends seen in $\mathcal{S}({\bf q})$}
%; hence it can only be detected on lattices where the Brillouin zone discretization contains these points. 

\newpage

\section*{Finite-size geometries for exact diagonalization}
\begin{figure}[t!]
\centering
\includegraphics[width=0.85\linewidth]{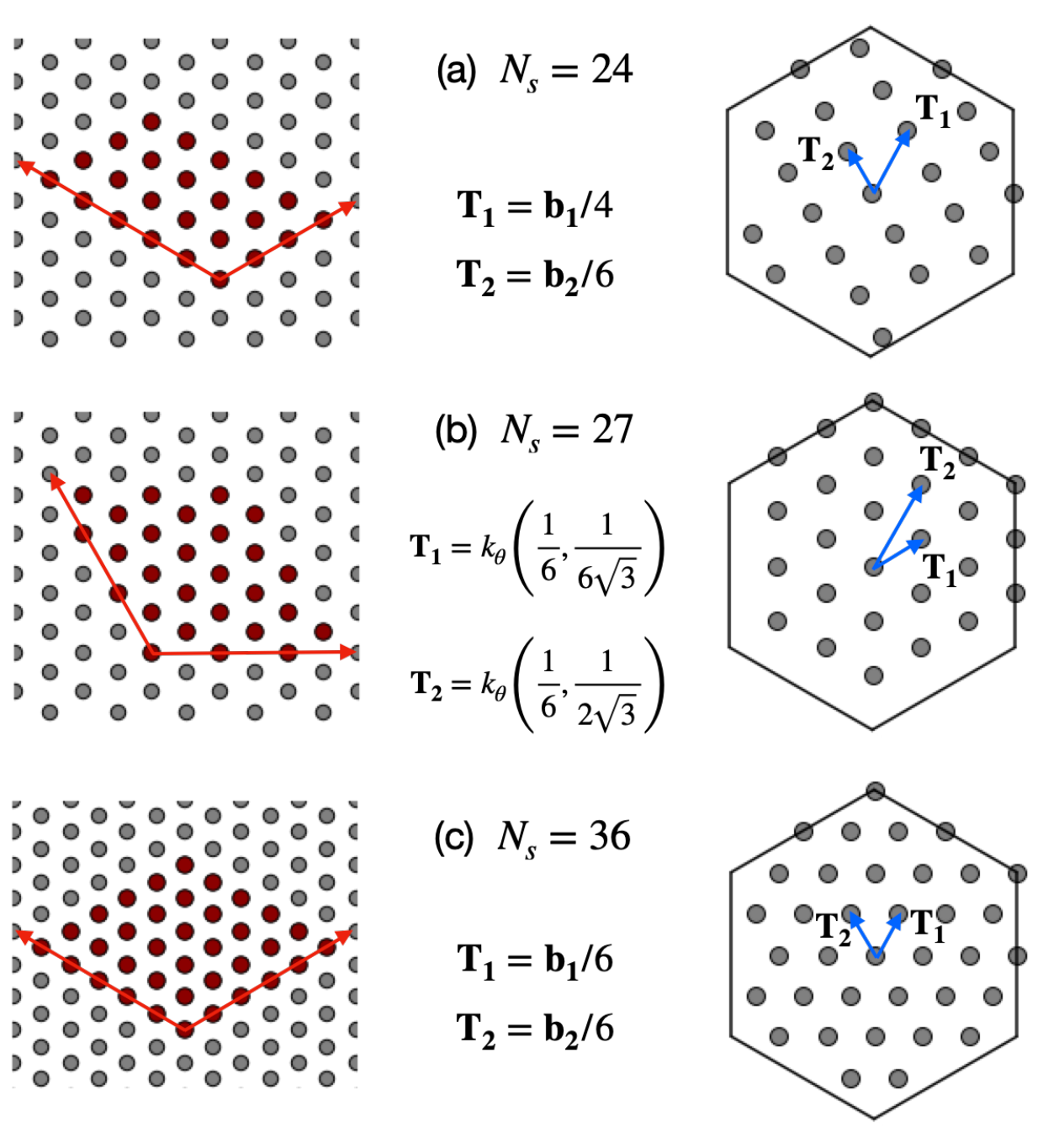}
\caption{Geometries used for exact diagonalization calculations in the main text. The plaquettes on the left indicate the number of moiré sites taken and the red arrows indicate the way in which periodic boundary conditions are applied. Each finite geometry determines a discretization of the Brillouin zone, with generators ${\bm T}_i$, that is shown on the right for each of the three sizes used.}
\label{fig:Lattice_Geometries}
\end{figure}

Details about the different finite size geometries used for ED calculations are presented in Fig. \ref{fig:Lattice_Geometries}. We used moiré reciprocal lattice vectors
\begin{align}
    {\bm b}_1=k_{\theta}\left(\frac{1}{2},\frac{\sqrt{3}}{2} \right)\quad\quad {\bm b}_2=k_{\theta}\left(-\frac{1}{2},\frac{\sqrt{3}}{2} \right),
\end{align}
with $k_{\theta}=4\pi/(\sqrt{3}a_M)$. We define different plaquettes in real space and apply periodic boundary conditions, which in turn determine a particular discretization of the moiré Brillouin zone, as illustrated in Fig. \ref{fig:Lattice_Geometries}. In particular our three geometries correspond to the geometries labelled 24C, 27A and 36, respectively in \cite{Lauchli_FCI}. Note that for the larger system sizes, $N_s=27\,,36$, the corresponding discretization of the Brillouin zone contains the ${\bf k}$ and ${\bf k}^{\prime}$ points, whose presence is important to determine the CDW phase.

\end{document}